\documentclass[prl, superscriptaddress, floatfix,twocolumn,showpacs,amsfonts,amsmath,amssymb,final]{revtex4}

\usepackage{graphicx}

\usepackage{amsmath}

\usepackage{amsfonts}

\usepackage{amssymb}

\usepackage{dcolumn} 

\usepackage{bm} 

\usepackage{color}

\usepackage{latexsym}
\begin{document}

\title{Fermi Surface Reconstruction by Dynamic Magnetic Fluctuations}

\author{Michael Holt}

\affiliation{School of Physics, University of New South Wales, Kensington 2052, Sydney NSW, Australia}

\author{Jaan Oitmaa}

\affiliation{School of Physics, University of New South Wales, Kensington 2052, Sydney NSW, Australia}

\author{Wei Chen}

\affiliation{Max-Planck-Institut f$\ddot{u}$r Festk$\ddot{o}$rperforschung, Heisenbergstrasse 1, D-70569 Stuttgart, Germany}

\author{Oleg P. Sushkov}

\affiliation{School of Physics, University of New South Wales, Kensington 2052, Sydney NSW, Australia}

\date{\rm\today}

\begin{abstract}

{We demonstrate that nearly critical quantum magnetic  fluctuations in strongly correlated electron
systems can change the Fermi surface topology and also lead to spin charge 
separation (SCS) in two dimensions.  To demonstrate these effects we consider 
a small number of holes injected into the bilayer antiferromagnet.
The system has a quantum critical point (QCP)  which separates  magnetically 
ordered and  disordered phases. We demonstrate that in the physically 
interesting regime there is a magnetically driven Lifshitz  point (LP) 
inside the magnetically disordered phase.   At the LP the topology of 
the hole Fermi surface is changed. We also demonstrate that in this regime 
the hole spin and charge necessarily separate when approaching  the QCP.
The considered model sheds light on generic problems concerning the physics of the cuprates.}

\end{abstract}






\maketitle
It is well known that static magnetic order in a conductor can influence the  Fermi surface (FS) topology due to the electron/hole diffraction from the ordered moments (Fermi surface reconstruction).  It is also well known that spin and charge are separate in one dimensional systems \cite{Tomonaga,Luttinger}. In the present work we address two generic problems - (1) Can {\it dynamic} magnetic fluctuations drive a change in the FS topology; (2) Is it possible to separate spin and charge in a two-dimensional (2D) system, and  if so what is the meaning of the separation. We give positive answers to both questions and demonstrate that these two problems are remarkably related. To address the generic problems we consider a specific model of a small number  of holes injected into the bilayer antiferromagnet (AF) with magnetic fluctuations driven by the interlayer coupling. We show that indeed purely dynamic short range AF correlations in the absence of a static AF order can cause a LP in which the topology of the FS is changed.  A similar model has been analyzed previously by Vojta and Becker \cite{Vojta99} where they also observed a LP. However in their case the LP was always in the AF ordered phase and therefore the central issue of the magnetic fluctuation driven LP has not been addressed. We also demonstrate that when the LP is in the disordered phase  the hole spin and charge necessarily separate when approaching the magnetic QCP. The possibility of spin-charge separation (SCS) in 2D has been discussed  previously within the context of the slave-boson method. The method  applied to the $t-J$ model implies the  separation ad hoc \cite{WenLee}. In the present analysis we do not make any ad hoc assumptions. The precise meaning of the  separation following from our analysis is different from that of the slave-boson method. 

Our interest in the problems is motivated by the cuprates, which, in our opinion, manifest both the LP and SCS. Below we explain the connection with cuprates. A reader who is not interested in cuprates can go directly to Eq.(1) where we start the analysis.

Lying at the center of the debate of high-$T_c$ superconductivity is whether it originates from a Fermi liquid or from a Mott insulator.  The angle-resolved photoemission spectroscopy (ARPES) indicates that the transition from a small to a large FS occurs in the hole doping range $0.1<x<0.15$~\cite{Hossain08, He11, Yang11}. Existence of hole pockets is consistent with the picture of dilute holes dressed by spin fluctuations, based on doping a Mott insulator~\cite{Liu92}. On the other hand, ARPES in the optimally to overdoped cuprates show  a large FS as expected by the Fermi liquid approach. This implies that there is a topological Lifshitz point~\cite{Lifshitz} (LP)  in the doping range $0.1<x<0.15$, where the Fermi surface changes from small to large. A phenomenological description of the LP based on the Fermi liquid picture was suggested in Ref.~\cite{Yang06}. DMFT calculations~\cite{Stanescu} with the Hubbard model also indicate the LP.

Magnetic quantum oscillation (MQO) in underdoped YBa$_{2}$Cu$_{3}$O$_{y}$ \cite{DoironLeyraud07} have revealed a small FS pocket, in contrast to the large FS observed on the overdoped side \cite{Vignolle08}. This again indicates the existence of a topological LP. The MQO measurements were performed in very strong magnetic fields, up to 80T. Therefore, a possible point of view is that the field induces a static magnetic structure and the structure causes the small Fermi surface reconstruction \cite{Harrison09}. However,  the small FS was observed in MQO up to 12\% doping \cite{Singleton10}, and it is unlikely that even an 80T field can generate a static AF order at such high doping. On the other hand the short range dynamic AF correlations always exist in the cuprates. Moreover, recent RIXS measurements \cite{LeTacon11} demonstrate remarkably that such correlations are practically doping independent, from Mott insulator to optimal doping. Based on this data one can conjecture that the cuprates are always close to magnetic criticality. This motivates us to study if the LP can be driven by short range, purely dynamic AF correlations. We consider a bilayer model for the sake of performing a controlled calculation. However, we believe that conceptually our conclusions are equally applicable to both single and multi-layer cuprates.

Now we turn to the discussion of  SCS and its relation to magnetic quantum criticality. Optimally and overdoped cuprates do not have any static magnetic order. On the other hand, the underdoped cuprates possess a static incommensurate magnetic order at zero temperature. A magnetic QCP separating these two regions was predicted in \cite{Milstein08}. In La$_{2-x}$Sr$_x$CuO$_4$ the QCP is  smeared out because of disorder. However, in YBa$_{2}$Cu$_{3}$O$_{y}$ the QCP is located experimentally at doping  $x \approx 0.09$ ($y\approx 6.47$) 
\cite{stock08, Hinkov08, Haug10}. In the magnetically ordered phase the hole does not carry usual spin, instead it carries only a psudospin which marks the sublattice. The pseudospin interacts with a magnetic field in a very  unusual way \cite{Braz89,Ramaz08} and this is the meaning of the partial SCS in the magnetically ordered phase \cite{Milstein08}. On the other side of the QCP there is no static magnetic order, and hence spin and charge  are united. In the present work we analyse the process of SCS at the QCP. The model considered here has 
only commensurate magnetic ordering, so we put aside incommensurability in the cuprates. 

We consider  the  $t-t^{\prime}-t^{\prime\prime}-J$ model defined by the following Hamiltonian on the bilayer square lattice
\begin{eqnarray}
\label{eq:tJhamiltonian}
H &=& J\sum_{\langle i,j\rangle}({\bf S}_i^{(1)}\cdot {\bf S}_j^{(1)} + {\bf S}_i^{(2)}\cdot {\bf S}_j^{(2)}) + J_{\perp}\sum_{i}{\bf S}_i^{(1)}\cdot {\bf S}_i^{(2)} \nonumber\\
&-& \sum_{\langle i,j \rangle}t_{i,j}(c_{i\sigma}^{\dagger} c_{j\sigma} + c_{j\sigma}^{\dagger} c_{i\sigma}) \ ,
\end{eqnarray}
\vspace{-3pt}
\noindent where $c_{i\sigma}^{\dagger}$ is the creation operator of an electron with spin  $\sigma=\uparrow,\downarrow$  at site $i$ on the top plane, ${\bm S}_i^{(1)}=\frac{1}{2}c_{i\mu}^{\dagger}{\bm \sigma}_{\mu\nu}c_{i\nu}$, and $t_{i,j}=\left\{t, t^{\prime}, t^{\prime\prime}\right\}$ is the hopping integral between nearest, next-nearest, and next-next nearest neighbour sites respectively. The superscripts (1), (2) indicate the layers. Hereafter we set $J = 1$. A no-double-occupancy constraint is imposed. It is well known that without holes (half-filling) the model has an O(3) magnetic QCP at $J_{\perp} \approx 2.5$ \cite{Sandvik95} separating the magnetically disordered and the AF ordered phases. The hopping integrals $t,t',t''$ result in charge dynamics if a hole is injected into the system. The longer range hopping integrals $t',t''$ are crucial as we will explain later. Note that we consider the zero temperature case, so the magnetic ordering in the AF phase is consistent with the Mermin-Wagner theorem.

We focus on small doping, $x \ll 1$, such that it does not influence the magnetic fluctuations. Magnetic fluctuations and the QCP are driven by  the interlayer coupling $J_{\perp}$. The holes  fill the rigid band formed by the magnetic quantum fluctuations.  To address the problems formulated above it is sufficient to calculate the single hole Green's function. Certainly at sufficiently high concentration of holes they start to influence the magnetic fluctuations and hence the rigid band approach fails. However, we do not need to go to such high concentrations to draw our conclusions. Such an approach is only possible because the magnetic dynamics are driven by $J_{\perp}$ and are independent of the hole concentration. This is a significant simplification compared to the $t-J$/Hubbard model, where doping is the only ``handle''.

We find that close to the $O(3)$ QCP, the hole dispersion has minima at ${\bf k}=(\pm \frac{\pi}{2},\pm \frac{\pi}{2})$. This results in  small hole pockets  similar to that in the cuprates at small doping. The pockets are formed due to strong in-plane AF correlations which diminish the nearest site hopping $t$. Upon increasing $J_{\perp}$ the in-plane AF correlations are reduced and the dispersion minima gradually shift and reach ${\bf k}=(\pm\pi,\pm\pi)$ at some value  $ J_{\perp}^{LP}$.  This is the position of topological LP at $x\to 0$. The in-plane AF correlations are diminished at  $J_{\perp}\to \infty$ and hence the hole dispersion is almost like that in the normal Fermi liquid, but a factor of two reduced due to $J_{\perp}$.
\begin{eqnarray}
\label{e0}
\epsilon_{k}^{(0)} &=& 2t\gamma_{k} + 2t'\gamma_{k}^{'} + 2t''\gamma_{k}^{''}\\
\gamma_{k} &= &\frac{1}{2}[\cos k_{x} + \cos k_y]\nonumber\\
\gamma_{k}^{'} &=& \cos k_x\cos k_y, \ \ \ \gamma_{k}^{''} = \frac{1}{2}[\cos(2k_{x}) + \cos(2k_{y})]\nonumber
\end{eqnarray}
At  $J_{\perp}< J_{\perp}^{LP}$ there are  four FS pockets and at $J_{\perp}> J_{\perp}^{LP}$ there is only one pocket centered at $(\pi,\pi)$ according to (\ref{e0}).  It is important to make it clear that as soon as we consider only one hole, we observe just a crossover from one shape of the dispersion to another. However, at a finite concentration of holes this results in a sharp LP
with MQO  frequency  jump by a factor of four. 

Magnetic excitations in the magnetically disordered phase are triplons. To describe the triplons we employ the spin-bond operator mean field technique \cite{Yu99}. This technique gives a QCP at  $J_{\perp}^{c}\approx 2.31$ which is reasonably close to the value $2.525$ known from Quantum Monte Carlo \cite{Sandvik95}. All necessary Eqs. describing the triplon dynamics have been derived in Refs. \cite{Yu99}. One can certainly employ a more accurate  Brueckner technique \cite{Kotov98}. However, this technique  is more involved while the bond operator mean field approach has sufficient accuracy for our purposes and we chose it for simplicity.

To describe the hole dressed by triplons we use the self-consistent Born approximation (SCBA) which disregards vertex corrections. This approximation has been widely used to study hole dynamics in the AF background \cite{Liu92}. In the AF background the single loop vertex correction is zero due to kinematic constraints and hence the SCBA is very accurate. In the present case of the magnetically disordered background the single loop vertex correction is  nonzero. However, the correction is suppressed by the parameter $1/N$, where $N = 3$ is the number of the triplon components \cite{Sushkov00}. To confirm the accuracy of the SCBA we compare results with that of numerically exact dimer series expansions. Note that here we are working in terms of the true spin of the hole, while in the case of the AF background one has to work in terms of pseudospin.

The  SCBA results in the following Dyson's equation for hole Green's function
\begin{eqnarray}
\label{eq:fullholegreen}
G({\bf k},\epsilon) = \bigg (\epsilon - \epsilon_{\bf k}^{(0)}  - 3\sum_{q} g_{{\bf k - q,q}}^{2} G({\bf k - q},\epsilon - \Omega_{q})  + i0 \bigg )^{-1} 
\nonumber
\end{eqnarray}
\noindent where the factor $3$ comes from the three different polarizations of the intermediate triplon. The bare hole dispersion  $\epsilon_{k}^{(0)}$ is given by (\ref{e0}). The hole-triplon vertex is $g_{k,q} =-\frac{1}{\sqrt{N}}\left[u_{q}\Gamma_{k}+ v_{q}\Gamma_{k+q}\right] -\frac{J}{\sqrt{N}}\gamma_{q}(u_{q} + v_{q})$, where $\Gamma_{p}=[2t\gamma_{p}+2t'\gamma_{p}'+2t''\gamma_{p}'']$, and $u_{q}$ and $v_{q}$ are the triplon Bogoliubov coefficients. These coefficients as well as the triplon energy  $\Omega_{q}$ were calculated  in \cite{Yu99}. Note, that on approaching the QCP the vertex diverges at ${\bf q} \to (\pi, \pi)$ and this leads to the SCS as we discuss below. A similar Dyson's equation  was used in Ref. \cite{Brunger06},  but with a different vertex $g_{k,q}$. We believe that the vertex in \cite{Brunger06} is wrong. We solve Eq. \eqref{eq:fullholegreen} numerically on a $128\times 128$ cluster with energy resolution $\Delta\epsilon=0.02$. The quasiparticle dispersion is given by the position of the $\delta$-function peak in the hole spectral function  $A({\bf k},\epsilon) = -\frac{1}{\pi}Im[G({\bf k},\epsilon)]$.

\begin{figure}
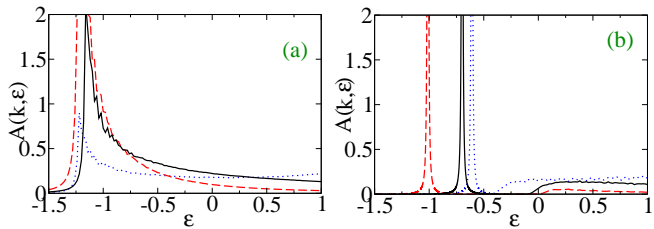

    \begin{center}
     \includegraphics[width=0.49\columnwidth,clip]{MD231A.eps}
     \includegraphics[width=0.49\columnwidth,clip]{MD300A.eps}
    \end{center}
\vspace{-15pt}
    \caption{(color online) The hole spectral functions at $t=0.5$, $t'=t''=0$, for $J_{\perp}=2.31$ $(a)$ and $J_{\perp}=3$ $(b)$. Here we show  ${\bf k}=(\frac{\pi}{2},\frac{\pi}{2})$ (black, solid), $(\pi,\pi)$ (red, dashed), and $(0,0)$ (blue, dotted). There is a significant distant incoherent part, the quasiparticle residues for $J_{\perp}=3$ are $Z_{(\frac{\pi}{2},\frac{\pi}{2})}=0.66$, $Z_{(\pi,\pi)}=0.88$,  $Z_{(0,0)}=0.32$. }
\label{0500spectra}
\end{figure}

We start from the case of small hopping, $t=0.5$, $t'=t''=0$. We plot in Fig.\ref{0500spectra} the spectral functions at values $J_{\perp}=2.31$ and $J_{\perp}=3$. Note that $J_{\perp}=2.31$ is exactly the position of the QCP obtained from the mean field triplon analysis. Spectra in Fig.\ref{0500spectra}(a) do not show any  quasiparticle peaks, instead there are only power cuts. This is similar to the Green's function of an immobile magnetic impurity at the QCP~\cite{Sushkov00,Vojta00}.  Remarkably the hole mobility does not influence this behaviour. The power cuts imply that the spin is distributed around the hole in a diverging cloud indicating SCS at the QCP, see discussion below. On the other hand spectra in Fig. \ref{0500spectra}(b)  show quasiparticle peaks separated by the triplon gap $\Delta$ from the incoherent spectra. Figure \ref{0500spectra}(b) shows the dispersion minimum at  ${\bf k} = (\pi,\pi)$, while the cut position in Fig. \ref{0500spectra}(a) is practically the same for all momenta. Hence, we conclude that the position of the LP in this case coincides with  that of the QCP. 

We also perform dimer series expansion calculations for $t=0.5$, $t'=t''=0$, and compare the results with the SCBA. 
The series expansion method allows one to determine only the quasiparticle dispersion. Naturally the method does not 
converge close to the QCP, as there are no quasiparticles there. However, at $J_{\perp}=3$ the method works well and agreement  
between the SCBA and series is good, for example the SCBA band width is $\epsilon_{(0,0)}-\epsilon_{(\pi,\pi)}=0.40$ and the series 
band width is $0.41$.

\begin{figure}
    \begin{center}
\includegraphics[width=0.49\columnwidth,clip]{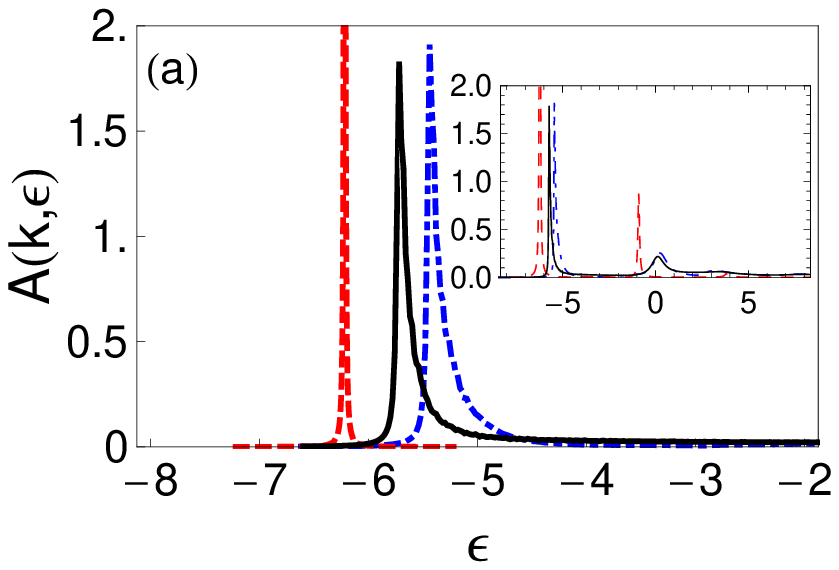}
\includegraphics[width=0.49\columnwidth,clip]{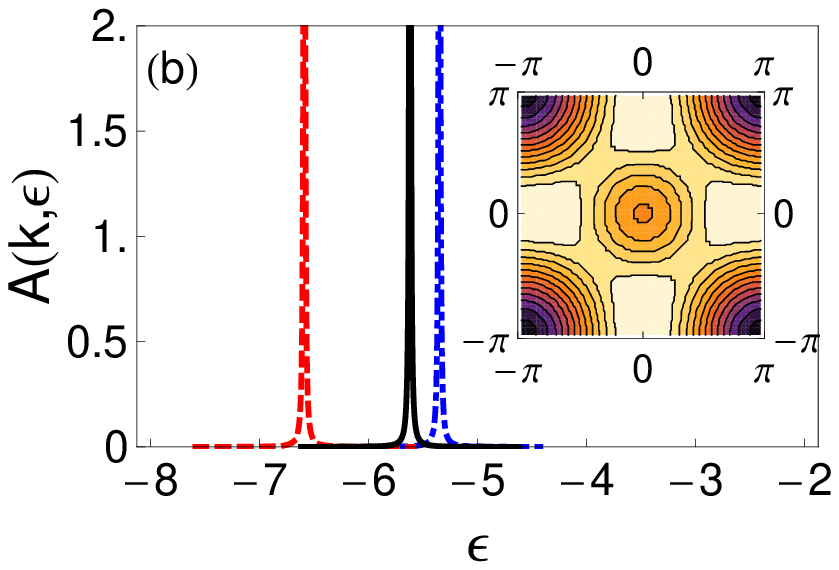}
    \end{center}
\vspace{-22pt}
    \caption{(color online) The hole spectral function at $t = 3.1$, $t'=t''= 0$, for $J_{\perp}=2.31$ $(a)$ and $J_{\perp}=3$ $(b)$. Values of the momentum are ${\bf k}=(\frac{\pi}{2},\frac{\pi}{2})$ (black, solid), $(\pi,\pi)$ (red, dashed), and $(\pi,0)$ (blue, dotted). The quasiparticle residues for $J_{\perp}=3$ are $Z_{(\frac{\pi}{2},\frac{\pi}{2})}=0.31$, $Z_{(\pi,\pi)}=0.80$,  $Z_{(\pi,0)}=0.35$. The inset in Fig.(a) shows spectral functions in the broader energy range and  the inset in Fig.(b) shows the map of the hole dispersion. The dark region is minimum of the dispersion.}
    \label{fig:spectralfermit310tp00tpp00}
\end{figure}

In the strong coupling limit, $t=3.1$, we rely on the SCBA since the series expansion does not converge. Spectral functions for $t=3.1$, $t'=t''=0$ are shown in Fig. \ref{fig:spectralfermit310tp00tpp00} for $J_{\perp}=2.31$ and $J_{\perp}=3.00$. In this case there is no LP in the disordered phase, as the bottom of the band is always at ${\bm k}=(\pi,\pi)$. Hence the LP is inside the magnetically ordered phase in agreement with \cite{Vojta99}. At the band bottom there  are well defined quasiparticles even at the QCP.
This indicates that there is no SCS. The spectra at  ${\bm k}=(\frac{\pi}{2},\frac{\pi}{2})$ and ${\bm k}=(\pi,0)$  have cuts at the QCP. However, these are the high energy states  which are irrelevant at small doping.

\begin{figure}
    \begin{center}
     \includegraphics[width=0.49\columnwidth,clip]{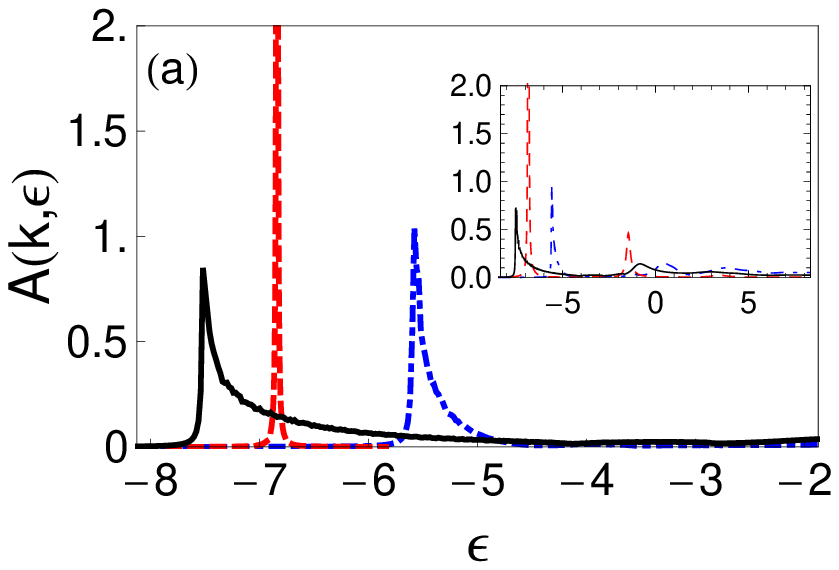}
     \includegraphics[width=0.49\columnwidth,clip]{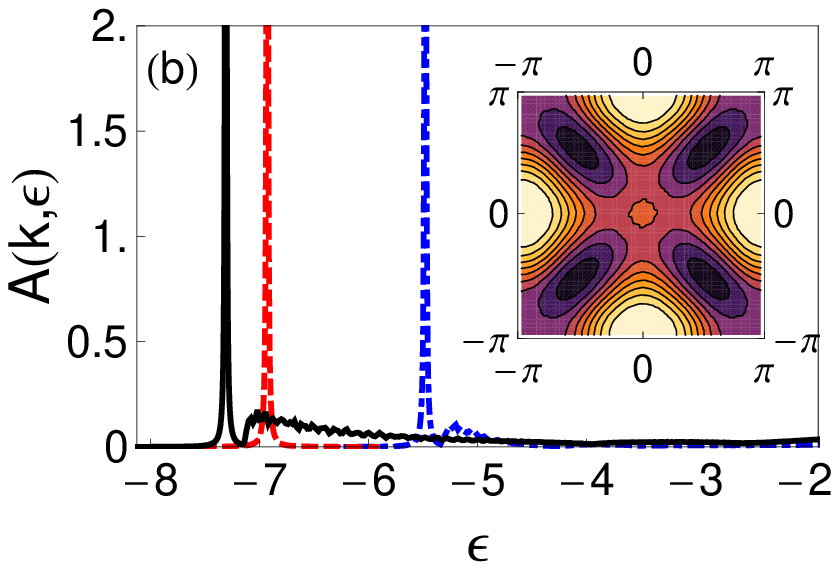}
     \includegraphics[width=0.49\columnwidth,clip]{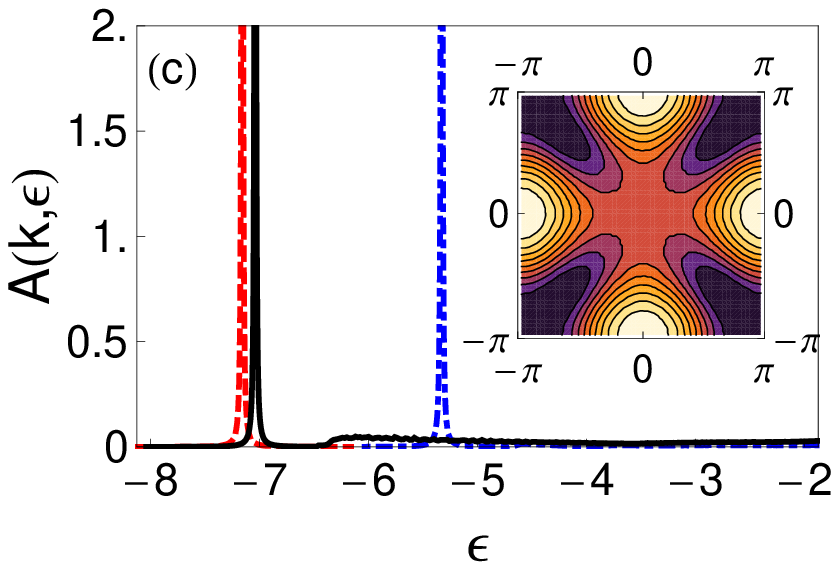}
     \includegraphics[width=0.49
\columnwidth,clip]{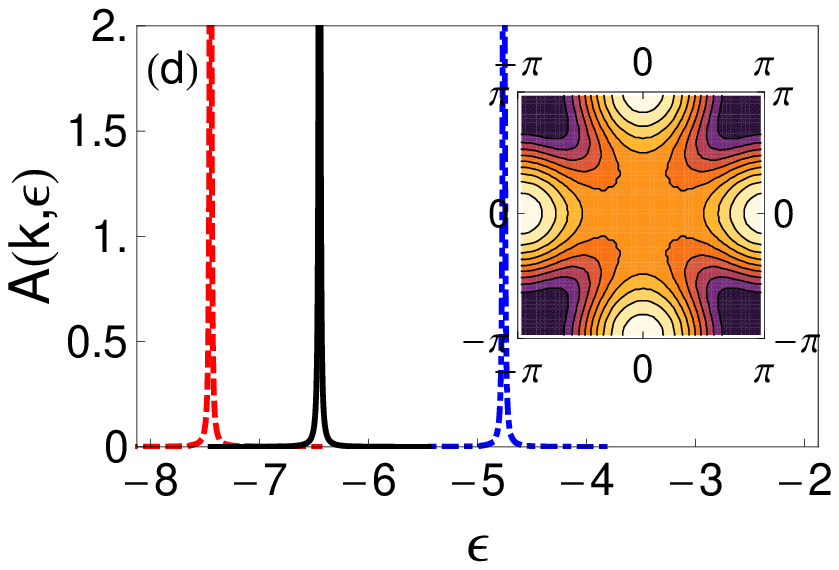}
    \end{center}
\vspace{-20pt}
    \caption{(color online) The hole spectral function at $t = 3.1$, $t' = -0.8$, $t'' = 0.7$ for  $J_{\perp} = 2.31$(a), $2.50$(b), $3.00$(c) and $4.00$(d).  Values of momentum are ${\bf k}=(\frac{\pi}{2},\frac{\pi}{2})$ (black, solid), $(\pi,\pi)$ (red, dashed), and $(\pi,0)$ (blue, dotted). The quasiparticle residues for $J_{\perp}=3$ are $Z_{(\frac{\pi}{2},\frac{\pi}{2})}=0.29$, $Z_{(\pi,\pi)}=0.77$,  $Z_{(\pi,0)}=0.20$; The inset in Fig.(a) shows spectral functions in the broader energy range and  the insets in Fig.(b)-(d) show  maps of the hole dispersion. The dark region is minimum of the dispersion.}    
\label{fig:spectralfermit310tp080tpp070}
\end{figure}

The last and the most important set of parameters, $t = 3.1$, $t' = -0.8$, $t'' = 0.7$, roughly corresponds to the parameters of the cuprates. Although we do not intend to simulate the cuprates quantitatively, we choose the parameters relevant to cuprates such that the interplay of various mechanisms can be compared on a similar energy scale. The spectral functions  are shown in  Fig. \ref{fig:spectralfermit310tp080tpp070} for several values of $J_{\perp}$. The dispersion maps shown in  the insets clearly demonstrate that the LP is located at $J_{\perp}\approx 3$ within the magnetically disordered phase. This topological transition is caused by fully dynamic antiferromagnetic correlations. The demonstration of the possibility of the fully dynamic scenario is the first major conclusion of the present work. Our analysis assumes small doping, practically we need $x < 0.1$ when the FS built on maps Fig.~3(b) and Fig.~3(c) 
are topologically different.
Why the longer range hoppings have qualitatively changed the situation? At $t'=t''=0$ the LP is in the ordered phase but already rather close to the QCP. A hopping $t''>0$ pushes the bare dispersion (\ref{e0}) at the nodal point ${\bm k}=(\pi/2,\pi/2)$ down helping magnetic fluctuations to form a small pocket. A pretty small positive $t''$ is sufficient to shift the LP to the disordered phase. The role of $t'$ is less important. The shift of the LP is due to the tuning of the longer range hoppings. The ``tuning'' has been performed by nature in the cuprates where the qualitative importance of $t',t''$ is well known. These parameters give asymmetry between the hole and the electron doping. Holes go to the nodal points while electrons go to the antinodal ones resulting in dramatically different Fermi surfaces and magnetic properties.
We follow nature and rely on the same mechanism.

Our second major conclusion follows from the first one and it concerns SCS at the QCP. According to Fig \ref{fig:spectralfermit310tp080tpp070}(a), at the QCP the lowest energy spectral function, ${\bf k}=(\frac{\pi}{2}, \frac{\pi}{2})$, does not have a pole, but only a cut. According to previous studies for an immobile impurity, a cut indicates that the spin density is distributed in a power cloud around the hole \cite{Sushkov00,Vojta00}. Having a similar Green's function we directly project the results of Refs. \cite{Sushkov00, Vojta00} to the present case. When approaching the QCP from the magnetically disordered phase the quasiparticle residue approaches zero, $Z \propto \Delta^z$, $z\approx 0.4$, as the triplon gap $\Delta$ approaches zero. The fraction of spin localized at  the hole  goes to zero $\propto Z$. The rest of spin is distributed around the hole over a disk of radius $R\propto 1/\Delta$. At $r \ll R$ the spin density is  $\propto 1/r^{\alpha}$, $1 < \alpha < 1.5$. Therefore the average radius of the spin cloud $\langle r\rangle \sim R$ diverges at the QCP, indicating SCS. On the other side of the QCP, deep inside the AF phase, the hole interaction with a magnetic field is described by pseudospin~\cite{Braz89,Ramaz08}.  This interaction implies a partial SCS \cite{Milstein08}.  Evolution of the spin cloud when approaching the QCP {\it from the AF phase} is not clear at present. Because of the diverging magnon cloud the hole effective mass also diverges at the QCP.  Drawing analogy with the cuprates we note that the effective mass measured in MQO \cite{Singleton10} diverges on approaching the QCP identified by neutron scattering \cite{stock08,Hinkov08,Haug10}.

{\it In conclusion}. We  consider the bilayer $t-t^{\prime}-t^{\prime\prime}-J$ model with strong interlayer coupling $J_{\perp}$. The hole doping is low, so the magnetic dynamics are driven only by the interlayer coupling. At  $J_{\perp} \to \infty$ the hole Fermi surface is connected and centered at ${\bf k}=(\pi,\pi)$. At certain $J_{\perp}^{LP}$ the AF correlations reconstruct the connected Fermi surface into four separate pockets centered close to  ${\bf k}=(\pm \frac{\pi}{2},\pm \frac{\pi}{2})$.  We have demonstrated that the LP where the topology of the FS is changed can be located within the magnetically disordered phase. The LP is driven by purely dynamic AF correlations in absence of any  static magnetic order.  The physics behind the LP is the following. Strong AF correlations make the nearest site hopping $t$ almost idle,  while the distant hoppings $t^{\prime}$ and $t^{\prime\prime}$ are influenced to a much lesser extent. A balance of these effects gives rise to the LP. Because of the importance of the AF correlations the LP is located not very far from the magnetic QCP.

We have also demonstrated that if the LP is located in the magnetically disordered phase then, on approaching the magnetic QCP, the hole spin and charge separate. The separation scale is equal to the magnetic correlation length which diverges at the QCP.

We are not aware of 2D materials with a low concentration of charge carriers and with a magnetic QCP driven by a separate parameter. Perhaps such materials will be synthesised in the future and/or the model can be realised with cold atoms. However, the most important outcome of the analysis is the demonstration of the principal possibility of the topological transition and spin-charge separation. We believe that conceptually the results are applicable to the single and multi-layer cuprates.

We are grateful to A. V. Chubukov and G. Khaliullin for useful comments. Numerical calculations performed in this work were done at the National Computational Infrastructure at the ANU, under project u66.

\end{document}